\documentclass[10pt]{article} 
\usepackage[preprint]{rlc}

\usepackage{amssymb}            
\usepackage{mathtools}          
\usepackage{mathrsfs}           
\mathtoolsset{showonlyrefs}     
\usepackage{graphicx}           
\usepackage{subcaption}         
\usepackage[space]{grffile}     
\usepackage{url}                

\usepackage{algorithm}
\usepackage{algpseudocode}
\usepackage{booktabs}

\newtheorem{theorem}{Theorem}

\title{MARL-LNS: \\Cooperative Multi-agent Reinforcement Learning \\ via Large Neighborhoods Search}

\author{Weizhe Chen, Sven Koenig, Bistra Dilkina  \\
    \{weizhech, skoenig, dilkina\}@usc.edu \\
    University of Souhthern California
    }

\begin{document}

\maketitle

\begin{abstract}
Cooperative multi-agent reinforcement learning (MARL) has been an increasingly important research topic in the last half-decade because of its great potential for real-world applications. Because of the curse of dimensionality, the popular "centralized training decentralized execution" framework requires a long time in training, yet still cannot converge efficiently. In this paper, we propose a general training framework, MARL-LNS, to algorithmically address these issues by training on alternating subsets of agents using existing deep MARL algorithms as low-level trainers, while not involving any additional parameters to be trained. Based on this framework, we provide three algorithm variants based on the framework: random large neighborhood search (RLNS), batch large neighborhood search (BLNS), and adaptive large neighborhood search (ALNS), which alternate the subsets of agents differently. We test our algorithms on both the StarCraft Multi-Agent Challenge and Google Research Football, showing that our algorithms can automatically reduce at least 10\% of training time while reaching the same final skill level as the original algorithm. 
\end{abstract}

\section{Introduction}
In recent years, multi-agent reinforcement learning (MARL) has split into cooperative multi-agent reinforcement learning and competitive multi-agent reinforcement learning. Competitive MARL has many theoretical guarantees following the previous studies in game theory, and has substantial success in domains like Poker \cite{brown2018superhuman}, and Diplomacy \cite{gray2020human}. On the other hand, cooperative multi-agent reinforcement learning focuses more on training a group of agents and making them coordinate with each other when everyone shares the same goal, and has also succeeded in many real-world applications like autonomous driving \cite{zhou2020smarts}, swarm control \cite{huttenrauch2017guided}, and traffic scheduling \cite{laurent2021flatland}. 

While previous research has shown that MARL can converge to a good policy that wins in a game or finishes some tasks in a given time, training the agent efficiently is always one of the problems that people are looking into. 
The difficulty of training MARL algorithms comes from training both the reinforcement learning algorithm and the complexity of multi-agent systems. As training a single agent reinforcement learning, the total time of MARL heavily depends on the sampling efficiency of the environment, which makes sampling from multiple environments run in parallel now a common practice to balance the sampling time and training time, and even recently rewriting environments with Jax \cite{flair2023jaxmarl}. 
After parallel sampling, the CPU sampling time has become 40\% of the total time, making GPU also an unneglectable part of the time used in the training process.


In this paper, we propose a new learning framework that reduces the time used in the training process of cooperative MARL without harming the performance of the final converged policy. We split the training into multiple iterations, which we called \textit{LNS iterations}, and we only consider the training introduced by a fixed group of agents in each iteration. We choose a subset of agents that are used in training for a certain number of training iterations using existing deep MARL algorithms to update the neural networks, and then alternate the subsets of agents so that we do not overfit the policy to one certain subgroup of agents while we need to control all of them. We call the group of agents we are considering the \textit{neighborhood}, and call our framework large neighborhood search (MARL-LNS), whose name comes from similar methods used in combinatorial optimization \cite{shaw1998using}. Since we have only modified the training at a high level, integrating MARL-LNS with any existing MARL algorithm is both straightforward and simple, and we choose to integrate it with MAPPO in this paper.
We provide a theoretical analysis that after multiple LNS iterations, the optimal action learned in this reduced joint action space could still hold the same convergence guarantee provided by the low-level MARL algorithm.

Based on our framework, we provide three simple yet powerful algorithms: random large neighborhood search (RLNS), batch large neighborhood search (BLNS), and adaptive large neighborhood search (ALNS). None of the three proposed have additional parameters that need to be trained, and ALNS do not even have additional hyperparameters that need to be tuned. To show the capability of our framework, our algorithms rely on random choices of which group of agents is used in training and do not include any hand-crafted or learned heuristic to select the neighborhood. 

We test our algorithms in many scenarios in the popular StarCraft Multi-Agent Challenge (SMAC) \cite{samvelyan19smac} and Google Research Football (GRF) \cite{kurach2020google} environments, and show that our simple algorithms can reach at least the same level of performance while we can be around $10\%$ faster than the baselines in terms of total wall clock time. We provide an ablation study on how the number of agents in the training neighborhood at each period, i.e., the neighborhood size, affects the performance of the algorithms and the training time. 

\section{Related Works}

Cooperative multi-agent reinforcement learning has been a popular research topic in the past few years. The most popular solution framework is the centralized training decentralized execution (CTDE) framework, in which two of the most popular lines of research are the value-based research like VDN \cite{sunehag2017value}, QMIX \cite{DBLP:conf/icml/RashidSWFFW18}, QTRAN \cite{son2019qtran}, and policy-based research like MADDPG \cite{lowe2017multi}, COMA \cite{foerster2018counterfactual} and MAPPO \cite{yu2021surprising}. 
A few recent works propose using only local neighborhood information at each timestep to keep the joint state-action space small \cite{zhang2022neighborhood}. Our algorithm framework is different from those by not changing the neighborhood used at each timestep in one episode, but only changing it after enough training iterations. Other works like VAST \cite{phanNeurIPS21} proposed to factorize the coordination into a sum over a few smaller sub-groups with smaller joint-action space, while our algorithm will only train one sub-group at one time. \citet{DBLP:conf/icml/IqbalWPBWS21} also proposed to factor the agents into random subpartitions in value-based MARL, and our work is different from theirs in training in more than two groups and working on policy-based MARL. DER \cite{DBLP:journals/corr/abs-2301-10574} also proposed to use part of the data in training, but they focused on training the policy efficiently in the beginning, while we are focusing on the overall efficiency and effectiveness.

The idea of large neighborhood search has been extensively used in combinatorial optimization problems since proposed \cite{shaw1998using}, such as vehicle routing \cite{ropke2006adaptive} and solving mixed integer linear programming \cite{munguia2018alternating,song2020general}. Similar ideas have also been used to help convergence in game theory by fixing a group of agents \cite{DBLP:conf/ijcai/NairTYPM03}. Recently, the same technique has been introduced to multi-agent path finding, where they fix the path of a group of agents and replan the path of other agents at each iteration \cite{li2021anytime,huang2022anytime,li2022mapf}. In these algorithms, the decisions of the variables/agents in the chosen neighborhood are updated such that this "move" results in a better overall solution to the optimization problem. In this paper, we propose a framework that introduces similar ideas into the MARL community.
In one special case of our algorithm, our algorithm becomes iterative learning, which has been widely used in equilibrium findings \cite{daskalakis2010learning, DBLP:conf/atal/WangGVTA18, chen2021temporal, DBLP:conf/ijcai/NairTYPM03}.
While their works mostly use iterations to make the training more stable and smooth, our work also focuses on reducing the time in each training iteration. 

While earlier works mostly focus on improving the speed in sampling on the CPU side \cite{flair2023jaxmarl}, and making parallel sampling a common practice, recently, there has been a growing interest in optimizing the overall training time, which takes GPU time into account. 
\citet{DBLP:conf/atal/YuYGCLLXHYWW23} proposed an asynchronous MARL based on MAPPO, thus reducing $10.07\%$ actual training time compared to MAPPO.
\citet{gogineni2023towards} explored neighbor a sampling strategy to improve cache locality and reached a $10.2\%$ end-to-end training time reduction.
\citet{DBLP:conf/nips/ChangM0PX22} proposed to guide agents with programs designed for parallelization, outperform the baselines in completion rate, and reduce the overall time by $13.43\%$.
Because the actual time reduction heavily relies on the specific configuration of the CPU and GPU used in training, the time reduction numbers are not comparable across different papers. Furthermore, these works that try to reduce the overall training time are orthogonal to each other, and ideally, all the methods, including the one that we will propose here, can be combined together to get a huge overall reduction.

\section{Preliminaries}
In multi-agent reinforcement learning, agents interact with the environment, which is formulated as a decentralized partially observable Markov decision process (Dec-POMDP). 
Formally, a Dec-POMDP is defined by a 7-tuple $\langle S, A, O, R, P, n, \gamma \rangle$. Here, $n$ is the total number of agents, $S$ is the state space, and $A$ is the action space. $O$ is the observation function. $P$ is the transition function. $R(s, a)$ denotes the shared reward function that every agent received. $\gamma$ is the discount factor used in calculating the cumulative reward. 
During the episode, agent $i$ use a policy $\pi_i(a_i |o_i)$ to produce an action $a_i$ from the local observation $o_i$. The environment starts with an initial state $s^1$, and stops when the environment provides a true in the \textit{done} signal. The objective is to find a policy $\pi= (\pi_1,\pi_2,\dots,\pi_n)$ within the valid policy space $\Pi$ that optimizes the discounted cumultative reward $\mathcal{J}=\mathbb{E}_{s^t,a^t \sim \pi} \sum_t \gamma^t \cdot r^t$, where $s^t$ is the state at timestep $t$, $r^t=R(s^t,a^t)$, and $a^t$ is the joint action at timestep $t$. 
A trajectory $\tau$ is a list of elements that contains all the information of what happens from the start to the end of one run of the environment, $\tau=<s^1,a^1,r^1,s^2,a^2,r^2, \dots, s^t,a^t,r^t>$. During the training of CTDE algorithms, the trajectories are split into trajectories in the view of every agent $\tau=(\tau_1,\tau_2,\dots,\tau_n)$, which contains useful information in terms of each agent and forms an individual trajectory. 
The information included in each $\tau_i$ depends on the algorithm that is used. 

\section{Large Neighborhood Search for MARL}

\subsection{Large Neighborhood Search Framework}


\begin{algorithm}[t]
   \caption{MARL-LNS: Large neighborhood search (LNS) framework used in this paper with MAPPO as low-level algorithm.}
   \label{alg:lns_2}
\begin{algorithmic}[1]
   \State Initialize value network $V$ and policy network $\pi$.
   \Repeat
   \State Choose the neighborhood $R=NeighborhoodSelect()$.
   \Repeat
   \State Reset the replay buffer
   \Repeat
   \State Sample trajectories $\tau = (\tau_1,\tau_2,\tau_3,\dots,\tau_n)$ from the environment according to $\pi$
   \State Save $\tau_{r_1},\tau_{r_2},\dots,\tau_{r_m}$ to the replay buffer, where $R= (r_1, r_2, \dots, r_m)$
   \Until{Sampled $Buffer\_length$ trajectories}
   \State Train $V$ and $\pi$ with the replay buffer using MAPPO
   \Until{Trained $N_{Training\_per\_neighborhood}$ rounds.}
   \Until{Has done $N_{LNS\_iterations}$ Iterations}
\end{algorithmic}
\end{algorithm}

Large neighborhood search is a popular meta-heuristic used in combinatorial optimization and multi-agent path finding for finding good solutions to challenging problems where finding the optimal solution is extremely time-consuming and infeasible in reality. Starting from an initial solution, part of the solution is selected, called a \textit{neighborhood}, and then destroyed. The optimizer then needs to rebuild the solution for the parts in the neighborhood while knowing the solution for the remaining parts and freezing their values for efficiency. 

While the high-level idea can lead to many completely different algorithms, we now give a detailed description of our framework in MARL, which specifies what destroying the neighborhood means, and what rebuilding the neighborhood means in the context of MARL. During training, we always keep a group of $m$ agents, and we call this group of agents a \textit{neighborhood} $R=\{r_1,r_2,\dots,r_m\}$, where $m$ is a hyperparameter specifying neighborhood size. Then, we sample a batch of data from the environment. During the sampling process, we decouple full information trajectory $\tau$ to $n$ subtrajectories $(\tau_1,\tau_2,\dots,\tau_n)$, where $\tau_i$ is to be used to train agent $i$ respectively. Then, we only keep the trajectories for the agents that are in the neighborhood, and put them into the replay buffer that is later used for training, i.e., we only save $\{\tau_{r_1}, \tau_{r_2}, \dots, \tau_{r_m}\}$ in the replay buffer rather than the whole trajectory $\tau$. 
We do one training when enough data for one batch is sampled, using existing training algorithms like MAPPO, and then clear the replay buffer. In this way, the training data only contains the information used to train the agents in the neighborhood. We repeat the trajectory sampling process several times until a new group of agents is resampled as the new neighborhood. We call the sampling of trajectories and training for one fixed neighborhood an \textit{LNS iteration}. We repeat a few LNS iterations until we meet our pre-set stopping criteria, which could be a total number of steps sampled, which is the same as other MARL algorithms.

Unlike many existing works focusing on sample efficiency, our algorithms gain efficiency by using fewer data for backpropagation in each batch of training. 
As a high-level approach, our algorithm is agnostic to how the data is used for training and how coordination is solved in the framework. Because of this, the lower-level MARL algorithm, which uses the data to train the neural network, is very flexible. 
In this paper, we choose to use MAPPO \cite{yu2021surprising} as the low-level algorithm as an example. In MAPPO, a centralized value function concatenates all observations from all agents in the environment. Even if some agents are not included in the neighborhood and thus not used in training, their observations are still kept in the input of the value function as global state information. 
Besides, the trajectory of each agent contains exactly the same state information as the original trajectories, and removing the trajectories outside the neighborhood saves us the space of copying the action information of those agents. We provide this algorithm framework in Alg.~\ref{alg:lns_2}, where lines 3, 8, 11, and 12 are the lines that are introduced because of our LNS framework.

Next, we theoretically show that the convergence guarantee will not be affected by the introduced framework, i.e., as long as the low-level algorithm can learn to cooperate, our framework will still be able to learn to do the same.
Our algorithm can be reduced to a block coordinate descent algorithm (BCD) by letting each variable used in the optimization be the policy of each agent, and the objective value is our reward function. BCD is studied a lot by optimization theory researchers \cite{tseng2001convergence, beck2013convergence, lu2015complexity}, and its convergence rate is proved under different conditions. Here, we provide a convergence guarantee that specifically proves that with MAPPO as the low-level algorithm, the expected cumulative reward of the learned policy from MARL-LNS is the same as the one from MAPPO:


\begin{theorem}
\label{thm:converge_inexact}
(Adapted from \cite{lyu2020convergence}) Assume the expected cumulative reward function $\mathcal{J}$ is continuously differentiable with Lipschitz gradient and convex in each neighborhood partition, and the training by the low-level algorithm guarantees that the training happening on the i-th neighborhood is bounded by a high-dimension vector $w_i$ on the joint policy space $\Pi$. Define the optimality gap as
$\Delta_i(\pi):=sup_{\hat{\pi} \in \Pi,|\hat{\pi}-\pi|\le c'w_i}J_{\hat{\pi}}-J_{\pi}$, where $c'$ is a constant.
Suppose $\sum_{i=1}^{\infty} w_i^2 < \infty$, and let the policy after the $k$-th LNS iteration be $\pi^k$. 
If the optimality gap is uniformly summable, i.e., $\sum_{i=1}^{\infty}\Delta_i < \infty$, then there exists some constant $c > 0$ such that for $i \ge 1$, 
    \begin{align}
    min_{1\le k \le i}  \sup_{\pi_0 \in \Pi}[-\inf_{\pi \in \Pi} \langle \nabla \mathcal{J}_{\pi^{k}}, \frac{\pi-\pi^{k}}{|\pi-\pi^{k}|}\rangle] \le \frac{c}{\sum_{k=1}^i w_k}    
    \end{align}
\end{theorem}

Specifically, the assumptions on $w_i$ are common assumptions for convergence in MARL, and are usually handled by the learning rate decay mechanism in the learning optimizer together with the clip mechanism in reinforcement learning algorithms like TRPO and PPO. Furthermore, because the expected reward function $\mathcal{J}$ is based on policy rather than action, the continuously differentiable condition is also satisfied in environments with a continuous policy space. This theorem guarantees that the convergence of MARL-LNS is irrelevant to the neighborhood size $m$ as well as what is included in each neighborhood. However, the learned policy could still be empirically worse than the policy learned by the low-level algorithm if the learning rate is not handled properly.

\subsection{Random Large Neighborhood Search}



\begin{algorithm}[t]
\small
   \caption{Neighborhoood selection function for Adaptive Large Neighborhood Search (ALNS).}
   \label{alg:alns}
\begin{algorithmic}[1]
    \State Initialize neighborhood size $m=2$.
    \Function{NeighborhoodSelect} {}
        \If {Agent performance was not improving in the last two LNS iterations}
            \State $m = min(m + 2^{\lfloor log_{2}^{m} \rfloor -1}, \lceil \frac{n}{2} \rceil)$
        \EndIf
        \State $R=random.choice(n,m)$
        \State \Return R
    \EndFunction
\end{algorithmic}
\end{algorithm}


While many previous works of large neighborhood search in CO and MAPF focus a lot on neighborhood selection, in this paper, we show the capability of our framework by choosing the neighborhood randomly and do not introduce any hand-crafted heuristics. We leave some discussion on how some simple heuristic-based approaches to neighborhood selection do not help the framework learn a better policy more efficiently in the appendix.
Specifically, the neighborhood selection part is instantiated with uniformly sample m agents from {1,..,n} without replacement as $random.choice$ do in NumPy. We call this algorithm the random large neighborhood search (RLNS).

\subsection{Batch Large Neighborhood Search}

While pure random can introduce a lot of variance to the training, here we also provide an alternative batch-based large neighborhood search (BLNS) algorithm, which differs from RLNS in the neighborhood selection function.
Unlike RLNS, before any training starts, we create one permutation $(p_1,p_2,\dots,p_n)$ of all agents. Again, for simplicity, the permutation is created randomly in this paper. After creating the permutation, we select the agents in order whenever we want to select the next group of neighborhoods. In other word, given a fixed  neighborhood size $m$, the first neighborhood would be $\{p_1,p_2,\dots,p_m\}$, the second would be $\{p_{m+1},p_{m+2},\dots,p_{2m}\}$, and keep going like this. If $m$ cannot be divided by $n$, the neighborhood that includes the last agent $p_n$ will also include agent $p_1$, and keep going from $p_2, p_3$ to $p_n$ again. 


\subsection{Adaptive Large Neighborhood Search}

While the RLNS and BLNS use a fixed neighborhood size, the adaptive large neighborhood size is recently becoming popular in the large neighborhood size community in combinatorial optimization and multi-agent path finding \cite{DBLP:journals/corr/abs-2107-10201, huang2022anytime}. Here, we propose another variant of MARL-LNS that adaptively changes the neighborhood size. In the beginning, we define a list of $k$ potential neighborhood size $M=[m_1,m_2,\dots,m_k]$, where $m_1 < m_2 < \dots < m_k$. In training, if in the last two LNS iterations, the evaluation performance is not getting any improvement, the current neighborhood size $m_i$ will be changed to $m_{i+1}$ in the next LNS iteration unless $m_i=m_k$ already. While this is orthogonal to the previously mentioned RLNS and BLNS, which focus on neighborhood selection, we combine this method with RLNS as a new algorithm, the adaptive large neighborhood search (ALNS). For implementations, because most environments have a smooth reward that encourages more agents to collaborate together, we can stay on a small neighborhood size most of the time. We advocate for setting $m_1=2$ and $m_i = \min(m_{i-1} + 2^{\lfloor \log_{2}(m_{i-1}) \rfloor -1}, \lceil \frac{n}{2} \rceil)$, where $n$ is the total number of agents, to ensure a gradual increase in neighborhood size, optimizing both efficiency and effectiveness. 
The corresponding ALNS pseudo-code is presented in Alg.~\ref{alg:alns}.

Besides the original benefit from MARL-LNS, the gradually growing neighborhood size $m$ gives an additional benefit that fits the nature of MARL: At the beginning of the training, both the value network and the policy network are far from accurate and optimal. In this period, MARL algorithms are mostly training value functions, and the neighborhood size does not affect the training of the value function. 
Later on, training on a subset of agents makes the training similar to iterative training, which reduces the size of the joint action space to speed up the convergence to local optimums. When it comes to the end of the training process, the neighborhood size will become large enough to cover the need of many agents in the environment to collaborate on a single task. 



\section{Experiments}

\subsection{Experimental Settings}

In this paper, we test our results on both StarCraft Multi-Agent Challenge (SMAC) and Google Research Football (GRF) environments. 
We use parameter sharing between agents because it has been shown to improve the training efficiency while not harming the performance \cite{DBLP:conf/icml/ChristianosPRA21}. We use some common practice tricks in MAPPO, including Generalized Advantage Estimation (GAE) \cite{schulman2015high} with advantage normalization, value clipping, and including both local-agent specific features and global features in the value function \cite{yu2021surprising}. 
For our algorithms, instead of setting a number of how many times of training is used for each LNS iteration, we use an equivalent version of providing the total number of different neighbors  $N_{Training\_per\_neighborhood}=N_{LNS\_iterations} / N_T$, where $N_T$ is the new hyperparameter we control. 
By default, the neighborhood size is half of the total number of agents. 
To show that our algorithm does not introduce extra fine-tuning effort, we do not change the hyperparameters used in our low-level algorithm MAPPO, e.g., the learning rate, the batch size, etc, as well as the network designs, and environment configurations. We provide more details in the appendix. This will affect the conclusion of Thm.~\ref{thm:converge_inexact}, but we will use our results to show that this is not affecting the effectiveness of our algorithm. For ALNS, we use the candidate neighborhood size list as we recommended. 

\subsection{SMAC Testbed}

We test our algorithms on 5 different random seeds. For each random seed, we evaluate our results following previous works: we compute the win rate over 32 evaluation games after each training iteration and take the median of the final ten evaluation win rates as the performance to alleviate the marginal distribution in this environment. We compare our algorithm with MAPPO, IPPO, and QMIX. We only test our results in scenarios classified as hard or super-hard, since many easy scenarios have been perfectly solved, and our algorithm will become iterative training, which has been studied a lot, in scenarios that include only 2 or 3 agents.

We report our results in Table.~\ref{tab:time_main} and Table.~\ref{tab:main}. In Table.~\ref{tab:time_main}, we observe that the time reductions are consistent across scenarios since the reduction comes from reducing the training data used. The time reduction is greater than that of other previous works on speeding up the overall time used in the training of MAPPO. Besides, comparing the reduction between the full training and the early number of steps results, we found that most savings are from the first half of the training, where the neighborhood size stays at a very small value, and the MARL algorithm is getting improvement on both the value network and the policy network given the huge space for improvement in this phase.
In Table.~\ref{tab:main}, the win rate of our algorithms is at least as good as the current algorithms while actually getting a higher final win rate in difficult scenarios like 5mvs6m and MMM2. This shows that our algorithm does not actually trade effectiveness in the trained policy for training efficiency, but gets the speedup without harming the performance. Furthermore, RLNS generally has a bigger variance in win rate than BLNS, which is coming from the pure-random-based neighborhood selection, but this randomness also enables RLNS to get a higher median win rate in the very hard 3s5zvs3s6z scenario where BLNS fails in 3 seeds and ends up with a low median value. 
When the neighborhood does not include both types of allies in 3s5zvs3s6z, the evaluation win rate drops quickly and needs a lot of extra training effort to make up for this drop. On the other hand, this scenario itself usually has a big marginal distribution, and both MAPPO, RLNS, and ALNS are still getting a policy with a win rate of less than $30\%$ in 2 out of the 5 seeds, leaving a great space for more stable policy training. Our ALNS is always one of the best algorithms given that it will at last use half of the total number of agents, but on the other hand, it never outperforms other algorithms, mostly because they are still in the same high-level framework. 

\begin{table}[htb]
\small
\centering
\begin{tabular}{@{}lllll@{}}
\toprule
 & RLNS \& BLNS & ALNS & ALNS (50\% Total Steps)& ALNS (70\% Total Steps)\\ \midrule
5mvs6m            & 5\%          &   5\%                        &  5\% & 5\%\\
MMM2              & 21\%         &   18\%  &             21\% & 19\%            \\
3s5zvs3s6z        & 8\%          &   12\%   &            15\% & 14\%            \\
27mvs30m          & 10\%         &   16\%    &       23\%   & 20\%             \\
10mvs11m          & 12\%         &   19\%  &        22\%     &   21\%         \\\bottomrule
\end{tabular}
\caption{Average total time reduction used by MARL-LNS compared to MAPPO in SMAC. The $50\%$ and $70\%$ Total steps indicate the scenarios in which each algorithm completes a respective portion of the total number of steps.}
\label{tab:time_main}
\end{table}

\begin{table*}[htb]
\small
\centering
\begin{tabular}{@{}llllllll@{}}
\toprule
           & MAPPO               & IPPO       & QMix                & RLNS (ours)               & BLNS (ours)          & ALNS (ours)     \\ \midrule
5mvs6m     & 89.1~(2.5)           & 87.5~(2.3)  & 75.8~(3.7)       & \textbf{96.9~(8.2)} & \textbf{96.9~(3.6)}  & \textbf{96.9~(3.6)}\\
MMM2       & 90.6~(2.8)           & 86.7~(7.3)  & 87.5~(2.6)         &   \textbf{96.9~(31.8)}& \textbf{96.9~(4.7)} &  \textbf{93.9~(2.4)}\\
3s5zvs3s6z & 84.4~(34.0) & 82.8~(19.1) & 82.8~(5.3)  & \textbf{87.5~(44.0)}          & 12.6~(31.8)         & \textbf{87.5~(34.7)}\\
27mvs30m   & \textbf{93.8~(2.4)}  & 69.5~(11.8) & 39.1~(9.8) & 90.6~(2.5)          & \textbf{93.8~(7.2)} & \textbf{93.8~(4.7)}\\
10mvs11m   & \textbf{96.9~(4.8)}  & 93.0~(7.4)  & 95.3~(1.0) & 93.8~(5.3) & \textbf{96.9~(2.4)}  & \textbf{96.9~(2.4)}\\ \bottomrule
\end{tabular}
\caption{Median evaluation win rate and standard deviation on SMAC testbed. }
\label{tab:main}
\end{table*}

\subsubsection{Ablation Study on Neighborhood Size}

After showing that our algorithms are good in terms of learning the policies, we now use an ablation study on neighborhood size to show how our method provides flexibility to trade off a tiny win rate for faster training time. 
As a special case, when changing the neighborhood size to 1, BLNS is iterative training, and when the neighborhood size is as big as the total number of agents, our framework is the same as the low-level algorithm MAPPO. We did not change the total environment steps because we do not see any benefit in doing that. 

We test BLNS on the 27mvs30m scenario, because it has the biggest number of agents. We show our results in Table.~\ref{tab:abla_m}. We observe that when increasing the neighborhood size, the final win rate is improving, while the time usage is also bigger. When we set the neighborhood size $m$ to $10$, the final performance is within one standard deviation of the low-level algorithm, while the training is $15\%$ faster. And if we set the neighborhood size $m$ to $5$, the final performance is within two standard deviations of MAPPO, with only $77\%$ of the original training time used. 
ALNS also achieves an average time saving between a neighborhood size of 5 and 10 and a final win rate as good as MAPPO, showing that it is a reasonably good algorithm that balances the performance of learned policy and training speed. Overall, ALNS and BLNS with $m=5$ are the two most dominant settings on the Pareto frontier of total training time and final win rate.
All the savings are because the sampling time in CPUs is only taking $44\%$ of the total training time of MAPPO, as shown in Table~\ref{tab:abla_m}. All other time is spent on transferring data between CPU and GPU and updating the neural networks on GPU, which is the time related to updating that can be largely saved by removing part of the training data. 
On the other hand, we can also observe that the standard deviation of the win rate of the policies from a small neighborhood size, i.e., $m=1$ or $m=3$, is growing bigger by the end of the training. This is because their policies have not actually converged, and given long enough time, their performance could also reach a good result. But understanding that the primary focus is training efficiency, they are not allowed to train any longer.

\begin{table*}[htb]
\small
\centering
\begin{tabular}{@{}llllllll@{}}
\toprule
m            & 1        & 3         & 5         & 10        & 15        & 27  (MAPPO)      & ALNS \\ \midrule
Win Rate  (\%)   & 3.1~(3.9) & 62.5~(28.3) & 87.5~(5.3) & 90.6~(8.2) & 93.8~(7.2) & 93.8~(3.8) & 93.8 (4.7)\\
Training Time (s) &     7.14     &    7.35       &     7.37      &     8.22      &     9.29      &   9.52   &    8.09\\
Updating Time (s) & 2.10 & 2.31 & 2.35 & 4.02 & 5.10 &  5.33 & 3.88\\\bottomrule
\end{tabular}
\caption{Median value and the standard deviation on evaluation win rate for MARL-RLNS with varied neighborhood sizes $m$, together with their average training time and average updating time (training time includes both sampling time and updating time with a little overlap) for 1k episodes on 27m\_vs\_30m scenario from SMAC. Specifically, when $m=27$, RLNS is the same as MAPPO.}
\label{tab:abla_m}
\end{table*}

\subsection{GRF Testbed}

We evaluate our results following the common practice in GRF: compared to the SMAC above, instead of evaluating in 32 evaluation games, the policies are evaluated in 100 rollouts, and instead of reporting the median value, the mean value is reported. Because most scenarios in GRF have less than 6 agents, we only test the algorithm in the corner scenario with 5 different random seeds.


Our results are shown in Table.~\ref{tab:grf}.  Even if this test case is naturally heterogeneous, we observe that giving a good hyperparameter to BLNS and RLNS will give our algorithm the same level of performance as MAPPO that is within one standard deviation, and applying the same group of hyperparameters to RLNS can learn a slightly worse policy with larger variance. In this hyperparameters setting, BLNS and RLNS are trained at least $14\%$ faster than MAPPO while ALNS is 25\% faster than MAPPO.
Additionally, GRF shows a more significant difference than SMAC when changing the number of different neighborhoods used in the training process. Medium size of 20 is enough for algorithms to explore collaborations with other agents while not changing it so regularly and introducing instability to the training.

\begin{table}[htb]
\small
\centering
\begin{tabular}{ccc}
\hline
                    & Win Rate (\%) & Time Reduction (\%)\\ \hline
BLNS ($m=7,N_T=20$) & 65.6(8.1)    &  15 \\
BLNS($m=5,N_T=20$)  & 57.4(6.0)    &  22\\
BLNS($m=7,N_T=10$)  & 42.4(3.1)    &  15\\
BLNS($m=7,N_T=5$)   & 42.2(1.8)    &  14\\ 
BLNS($m=7, N_T=40$) & 50.4(3.1)       & 15\\
RLNS($m=7,N_T=20$)  & 58.0(13.5)   &  15\\
ALNS($N_T=20$)      & 63.0(14.7)   & 25\\ 
MAPPO               & 65.53(2.19)  &  0 \\\hline
\end{tabular}
\caption{The average evaluation win rate of BLNS in different settings compared to RLNS and other baselines on GRF, together with their corresponding time reduction compared to MAPPO.}
\label{tab:grf}
\end{table}


\section{Conclusion}

In this paper, we propose a novel extensive neighborhood search framework (MARL-LNS) for cooperative MARL to train the policies more efficiently by using subgroups of agents, named neighborhood, during each training iteration. Building upon MARL-LNS, we design three algorithms RLNS, BLNS, and ALNS, that select the neighborhood differently. Specifically, our algorithm does not introduce any additional parameters to be trained, and ALNS is even hyperparameter-free. Through both theoretical and empirical analysis, we demonstrate that our algorithms significantly enhance training efficiency without compromising any training aspects, particularly in challenging environments such as MMM2 within SMAC, where they also contribute to learning a superior policy. 

\bibliography{main}

\begin{thebibliography}{40}
\providecommand{\natexlab}[1]{#1}
\providecommand{\url}[1]{\texttt{#1}}
\expandafter\ifx\csname urlstyle\endcsname\relax
  \providecommand{\doi}[1]{doi: #1}\else
  \providecommand{\doi}{doi: \begingroup \urlstyle{rm}\Url}\fi

\bibitem[Beck \& Tetruashvili(2013)Beck and Tetruashvili]{beck2013convergence}
Amir Beck and Luba Tetruashvili.
\newblock On the convergence of block coordinate descent type methods.
\newblock \emph{SIAM journal on Optimization}, 23\penalty0 (4):\penalty0
  2037--2060, 2013.

\bibitem[Brown \& Sandholm(2018)Brown and Sandholm]{brown2018superhuman}
Noam Brown and Tuomas Sandholm.
\newblock Superhuman ai for heads-up no-limit poker: Libratus beats top
  professionals.
\newblock \emph{Science}, 359\penalty0 (6374):\penalty0 418--424, 2018.

\bibitem[Chang et~al.(2022)Chang, Mu, Wu, Pan, and
  Xu]{DBLP:conf/nips/ChangM0PX22}
Can Chang, Ni~Mu, Jiajun Wu, Ling Pan, and Huazhe Xu.
\newblock {E-MAPP:} efficient multi-agent reinforcement learning with parallel
  program guidance.
\newblock In \emph{NeurIPS}, 2022.
\newblock URL
  \url{http://papers.nips.cc/paper\_files/paper/2022/hash/4f2accafe6fa355624f3ee42207cc7b8-Abstract-Conference.html}.

\bibitem[Chen et~al.(2021)Chen, Zhou, Wu, and Fang]{chen2021temporal}
Weizhe Chen, Zihan Zhou, Yi~Wu, and Fei Fang.
\newblock Temporal induced self-play for stochastic bayesian games.
\newblock \emph{arXiv preprint arXiv:2108.09444}, 2021.

\bibitem[Christianos et~al.(2021)Christianos, Papoudakis, Rahman, and
  Albrecht]{DBLP:conf/icml/ChristianosPRA21}
Filippos Christianos, Georgios Papoudakis, Arrasy Rahman, and Stefano~V.
  Albrecht.
\newblock Scaling multi-agent reinforcement learning with selective parameter
  sharing.
\newblock In Marina Meila and Tong Zhang (eds.), \emph{Proceedings of the 38th
  International Conference on Machine Learning, {ICML} 2021, 18-24 July 2021,
  Virtual Event}, volume 139 of \emph{Proceedings of Machine Learning
  Research}, pp.\  1989--1998. {PMLR}, 2021.
\newblock URL \url{http://proceedings.mlr.press/v139/christianos21a.html}.

\bibitem[Daskalakis et~al.(2010)Daskalakis, Frongillo, Papadimitriou,
  Pierrakos, and Valiant]{daskalakis2010learning}
Constantinos Daskalakis, Rafael~M Frongillo, Christos~H Papadimitriou, George
  Pierrakos, and Gregory Valiant.
\newblock On learning algorithms for nash equilibria.
\newblock In \emph{SAGT}, pp.\  114--125. Springer, 2010.

\bibitem[Foerster et~al.(2018)Foerster, Farquhar, Afouras, Nardelli, and
  Whiteson]{foerster2018counterfactual}
Jakob Foerster, Gregory Farquhar, Triantafyllos Afouras, Nantas Nardelli, and
  Shimon Whiteson.
\newblock Counterfactual multi-agent policy gradients.
\newblock In \emph{Proceedings of the AAAI conference on artificial
  intelligence}, volume~32, 2018.

\bibitem[Gogineni et~al.(2023)Gogineni, Wei, Lan, and
  Venkataramani]{gogineni2023towards}
Kailash Gogineni, Peng Wei, Tian Lan, and Guru~Prasadh Venkataramani.
\newblock Towards efficient multi-agent learning systems.
\newblock In \emph{Architecture and System Support for Transformer Models
  (ASSYST@ ISCA 2023)}, 2023.

\bibitem[Gray et~al.(2020)Gray, Lerer, Bakhtin, and Brown]{gray2020human}
Jonathan Gray, Adam Lerer, Anton Bakhtin, and Noam Brown.
\newblock Human-level performance in no-press diplomacy via equilibrium search.
\newblock In \emph{International Conference on Learning Representations}, 2020.

\bibitem[Hu et~al.(2023)Hu, Zhao, Zhou, and
  Li]{DBLP:journals/corr/abs-2301-10574}
Xunhan Hu, Jian Zhao, Wengang Zhou, and Houqiang Li.
\newblock Discriminative experience replay for efficient multi-agent
  reinforcement learning.
\newblock \emph{CoRR}, abs/2301.10574, 2023.
\newblock \doi{10.48550/arXiv.2301.10574}.
\newblock URL \url{https://doi.org/10.48550/arXiv.2301.10574}.

\bibitem[Huang et~al.(2022)Huang, Li, Koenig, and Dilkina]{huang2022anytime}
Taoan Huang, Jiaoyang Li, Sven Koenig, and Bistra Dilkina.
\newblock Anytime multi-agent path finding via machine learning-guided large
  neighborhood search.
\newblock In \emph{Proceedings of the AAAI Conference on Artificial
  Intelligence}, volume~36, pp.\  9368--9376, 2022.

\bibitem[H{\"u}ttenrauch et~al.(2017)H{\"u}ttenrauch, {\v{S}}o{\v{s}}i{\'c},
  and Neumann]{huttenrauch2017guided}
Maximilian H{\"u}ttenrauch, Adrian {\v{S}}o{\v{s}}i{\'c}, and Gerhard Neumann.
\newblock Guided deep reinforcement learning for swarm systems.
\newblock \emph{arXiv preprint arXiv:1709.06011}, 2017.

\bibitem[Iqbal et~al.(2021)Iqbal, de~Witt, Peng, Boehmer, Whiteson, and
  Sha]{DBLP:conf/icml/IqbalWPBWS21}
Shariq Iqbal, Christian A.~Schr{\"{o}}der de~Witt, Bei Peng, Wendelin Boehmer,
  Shimon Whiteson, and Fei Sha.
\newblock Randomized entity-wise factorization for multi-agent reinforcement
  learning.
\newblock In Marina Meila and Tong Zhang (eds.), \emph{Proceedings of the 38th
  International Conference on Machine Learning, {ICML} 2021, 18-24 July 2021,
  Virtual Event}, volume 139 of \emph{Proceedings of Machine Learning
  Research}, pp.\  4596--4606. {PMLR}, 2021.
\newblock URL \url{http://proceedings.mlr.press/v139/iqbal21a.html}.

\bibitem[Kurach et~al.(2020)Kurach, Raichuk, Sta{\'n}czyk, Zaj{\k{a}}c, Bachem,
  Espeholt, Riquelme, Vincent, Michalski, Bousquet, et~al.]{kurach2020google}
Karol Kurach, Anton Raichuk, Piotr Sta{\'n}czyk, Micha{\l} Zaj{\k{a}}c, Olivier
  Bachem, Lasse Espeholt, Carlos Riquelme, Damien Vincent, Marcin Michalski,
  Olivier Bousquet, et~al.
\newblock Google research football: A novel reinforcement learning environment.
\newblock In \emph{Proceedings of the AAAI Conference on Artificial
  Intelligence}, volume~34, pp.\  4501--4510, 2020.

\bibitem[Laurent et~al.(2021)Laurent, Schneider, Scheller, Watson, Li, Chen,
  Zheng, Chan, Makhnev, Svidchenko, et~al.]{laurent2021flatland}
Florian Laurent, Manuel Schneider, Christian Scheller, Jeremy Watson, Jiaoyang
  Li, Zhe Chen, Yi~Zheng, Shao-Hung Chan, Konstantin Makhnev, Oleg Svidchenko,
  et~al.
\newblock Flatland competition 2020: Mapf and marl for efficient train
  coordination on a grid world.
\newblock In \emph{NeurIPS 2020 Competition and Demonstration Track}, pp.\
  275--301. PMLR, 2021.

\bibitem[Li et~al.(2021)Li, Chen, Harabor, Stuckey, and Koenig]{li2021anytime}
Jiaoyang Li, Zhe Chen, Daniel Harabor, P~Stuckey, and Sven Koenig.
\newblock Anytime multi-agent path finding via large neighborhood search.
\newblock In \emph{Proceedings of the International Joint Conference on
  Artificial Intelligence (IJCAI)}, 2021.

\bibitem[Li et~al.(2022)Li, Chen, Harabor, Stuckey, and Koenig]{li2022mapf}
Jiaoyang Li, Zhe Chen, Daniel Harabor, Peter~J Stuckey, and Sven Koenig.
\newblock Mapf-lns2: fast repairing for multi-agent path finding via large
  neighborhood search.
\newblock In \emph{Proceedings of the AAAI Conference on Artificial
  Intelligence}, volume~36, pp.\  10256--10265, 2022.

\bibitem[Lowe et~al.(2017)Lowe, Wu, Tamar, Harb, Pieter~Abbeel, and
  Mordatch]{lowe2017multi}
Ryan Lowe, Yi~I Wu, Aviv Tamar, Jean Harb, OpenAI Pieter~Abbeel, and Igor
  Mordatch.
\newblock Multi-agent actor-critic for mixed cooperative-competitive
  environments.
\newblock \emph{Advances in neural information processing systems}, 30, 2017.

\bibitem[Lu \& Xiao(2015)Lu and Xiao]{lu2015complexity}
Zhaosong Lu and Lin Xiao.
\newblock On the complexity analysis of randomized block-coordinate descent
  methods.
\newblock \emph{Mathematical Programming}, 152:\penalty0 615--642, 2015.

\bibitem[Lyu(2020)]{lyu2020convergence}
Hanbaek Lyu.
\newblock Convergence and complexity of block coordinate descent with
  diminishing radius for nonconvex optimization.
\newblock \emph{arXiv preprint arXiv:2012.03503}, 2020.

\bibitem[Mungu{\'\i}a et~al.(2018)Mungu{\'\i}a, Ahmed, Bader, Nemhauser, and
  Shao]{munguia2018alternating}
Llu{\'\i}s-Miquel Mungu{\'\i}a, Shabbir Ahmed, David~A Bader, George~L
  Nemhauser, and Yufen Shao.
\newblock Alternating criteria search: a parallel large neighborhood search
  algorithm for mixed integer programs.
\newblock \emph{Computational Optimization and Applications}, 69\penalty0
  (1):\penalty0 1--24, 2018.

\bibitem[Nair et~al.(2003)Nair, Tambe, Yokoo, Pynadath, and
  Marsella]{DBLP:conf/ijcai/NairTYPM03}
Ranjit Nair, Milind Tambe, Makoto Yokoo, David~V. Pynadath, and Stacy Marsella.
\newblock Taming decentralized pomdps: Towards efficient policy computation for
  multiagent settings.
\newblock In Georg Gottlob and Toby Walsh (eds.), \emph{IJCAI-03, Proceedings
  of the Eighteenth International Joint Conference on Artificial Intelligence,
  Acapulco, Mexico, August 9-15, 2003}, pp.\  705--711. Morgan Kaufmann, 2003.
\newblock URL \url{http://ijcai.org/Proceedings/03/Papers/103.pdf}.

\bibitem[Phan et~al.(2021)Phan, Ritz, Belzner, Altmann, Gabor, and
  Linnhoff-Popien]{phanNeurIPS21}
Thomy Phan, Fabian Ritz, Lenz Belzner, Philipp Altmann, Thomas Gabor, and
  Claudia Linnhoff-Popien.
\newblock Vast: Value function factorization with variable agent sub-teams.
\newblock In \emph{Advances in Neural Information Processing Systems
  (NeurIPS)}, volume~34, pp.\  24018--24032. Curran Associates, Inc., 2021.

\bibitem[Rashid et~al.(2018)Rashid, Samvelyan, de~Witt, Farquhar, Foerster, and
  Whiteson]{DBLP:conf/icml/RashidSWFFW18}
Tabish Rashid, Mikayel Samvelyan, Christian~Schr{\"{o}}der de~Witt, Gregory
  Farquhar, Jakob~N. Foerster, and Shimon Whiteson.
\newblock {QMIX:} monotonic value function factorisation for deep multi-agent
  reinforcement learning.
\newblock In Jennifer~G. Dy and Andreas Krause (eds.), \emph{Proceedings of the
  35th International Conference on Machine Learning, {ICML} 2018,
  Stockholmsm{\"{a}}ssan, Stockholm, Sweden, July 10-15, 2018}, volume~80 of
  \emph{Proceedings of Machine Learning Research}, pp.\  4292--4301. {PMLR},
  2018.
\newblock URL \url{http://proceedings.mlr.press/v80/rashid18a.html}.

\bibitem[Ropke \& Pisinger(2006)Ropke and Pisinger]{ropke2006adaptive}
Stefan Ropke and David Pisinger.
\newblock An adaptive large neighborhood search heuristic for the pickup and
  delivery problem with time windows.
\newblock \emph{Transportation science}, 40\penalty0 (4):\penalty0 455--472,
  2006.

\bibitem[Rutherford et~al.(2023)Rutherford, Ellis, Gallici, Cook, Lupu,
  Ingvarsson, Willi, Khan, de~Witt, Souly, Bandyopadhyay, Samvelyan, Jiang,
  Lange, Whiteson, Lacerda, Hawes, Rocktaschel, Lu, and
  Foerster]{flair2023jaxmarl}
Alexander Rutherford, Benjamin Ellis, Matteo Gallici, Jonathan Cook, Andrei
  Lupu, Gardar Ingvarsson, Timon Willi, Akbir Khan, Christian~Schroeder
  de~Witt, Alexandra Souly, Saptarashmi Bandyopadhyay, Mikayel Samvelyan, Minqi
  Jiang, Robert~Tjarko Lange, Shimon Whiteson, Bruno Lacerda, Nick Hawes, Tim
  Rocktaschel, Chris Lu, and Jakob~Nicolaus Foerster.
\newblock Jaxmarl: Multi-agent rl environments in jax.
\newblock \emph{arXiv preprint arXiv:2311.10090}, 2023.

\bibitem[Samvelyan et~al.(2019)Samvelyan, Rashid, de~Witt, Farquhar, Nardelli,
  Rudner, Hung, Torr, Foerster, and Whiteson]{samvelyan19smac}
Mikayel Samvelyan, Tabish Rashid, Christian~Schroeder de~Witt, Gregory
  Farquhar, Nantas Nardelli, Tim G.~J. Rudner, Chia-Man Hung, Philiph H.~S.
  Torr, Jakob Foerster, and Shimon Whiteson.
\newblock {The} {StarCraft} {Multi}-{Agent} {Challenge}.
\newblock \emph{CoRR}, abs/1902.04043, 2019.

\bibitem[Schulman et~al.(2015)Schulman, Moritz, Levine, Jordan, and
  Abbeel]{schulman2015high}
John Schulman, Philipp Moritz, Sergey Levine, Michael Jordan, and Pieter
  Abbeel.
\newblock High-dimensional continuous control using generalized advantage
  estimation.
\newblock \emph{arXiv preprint arXiv:1506.02438}, 2015.

\bibitem[Shaw(1998)]{shaw1998using}
Paul Shaw.
\newblock Using constraint programming and local search methods to solve
  vehicle routing problems.
\newblock In \emph{Principles and Practice of Constraint Programming—CP98:
  4th International Conference, CP98 Pisa, Italy, October 26--30, 1998
  Proceedings 4}, pp.\  417--431. Springer, 1998.

\bibitem[Son et~al.(2019)Son, Kim, Kang, Hostallero, and Yi]{son2019qtran}
Kyunghwan Son, Daewoo Kim, Wan~Ju Kang, David~Earl Hostallero, and Yung Yi.
\newblock Qtran: Learning to factorize with transformation for cooperative
  multi-agent reinforcement learning.
\newblock In \emph{International conference on machine learning}, pp.\
  5887--5896. PMLR, 2019.

\bibitem[Song et~al.(2020)Song, Lanka, Yue, and Dilkina]{song2020general}
Jialin Song, Ravi Lanka, Yisong Yue, and Bistra Dilkina.
\newblock A general large neighborhood search framework for solving integer
  programs.
\newblock \emph{arXiv}, 2020.

\bibitem[Sonnerat et~al.(2021)Sonnerat, Wang, Ktena, Bartunov, and
  Nair]{DBLP:journals/corr/abs-2107-10201}
Nicolas Sonnerat, Pengming Wang, Ira Ktena, Sergey Bartunov, and Vinod Nair.
\newblock Learning a large neighborhood search algorithm for mixed integer
  programs.
\newblock \emph{CoRR}, abs/2107.10201, 2021.
\newblock URL \url{https://arxiv.org/abs/2107.10201}.

\bibitem[Sunehag et~al.(2017)Sunehag, Lever, Gruslys, Czarnecki, Zambaldi,
  Jaderberg, Lanctot, Sonnerat, Leibo, Tuyls, et~al.]{sunehag2017value}
Peter Sunehag, Guy Lever, Audrunas Gruslys, Wojciech~Marian Czarnecki, Vinicius
  Zambaldi, Max Jaderberg, Marc Lanctot, Nicolas Sonnerat, Joel~Z Leibo, Karl
  Tuyls, et~al.
\newblock Value-decomposition networks for cooperative multi-agent learning.
\newblock \emph{arXiv preprint arXiv:1706.05296}, 2017.

\bibitem[Tseng(2001)]{tseng2001convergence}
Paul Tseng.
\newblock Convergence of a block coordinate descent method for
  nondifferentiable minimization.
\newblock \emph{Journal of optimization theory and applications}, 109:\penalty0
  475--494, 2001.

\bibitem[Wang et~al.(2018)Wang, Guo, Vayanos, Tambe, and
  An]{DBLP:conf/atal/WangGVTA18}
Kai Wang, Qingyu Guo, Phebe Vayanos, Milind Tambe, and Bo~An.
\newblock Equilibrium refinement in security games with arbitrary scheduling
  constraints.
\newblock In Elisabeth Andr{\'{e}}, Sven Koenig, Mehdi Dastani, and Gita
  Sukthankar (eds.), \emph{Proceedings of the 17th International Conference on
  Autonomous Agents and MultiAgent Systems, {AAMAS} 2018, Stockholm, Sweden,
  July 10-15, 2018}, pp.\  919--927. International Foundation for Autonomous
  Agents and Multiagent Systems Richland, SC, {USA} / {ACM}, 2018.
\newblock URL \url{http://dl.acm.org/citation.cfm?id=3237836}.

\bibitem[Wu et~al.(2021)Wu, Yu, Ye, Zhang, Zhuo, et~al.]{wu2021coordinated}
Zifan Wu, Chao Yu, Deheng Ye, Junge Zhang, Hankz~Hankui Zhuo, et~al.
\newblock Coordinated proximal policy optimization.
\newblock \emph{Advances in Neural Information Processing Systems},
  34:\penalty0 26437--26448, 2021.

\bibitem[Yu et~al.(2021)Yu, Velu, Vinitsky, Wang, Bayen, and
  Wu]{yu2021surprising}
Chao Yu, Akash Velu, Eugene Vinitsky, Yu~Wang, Alexandre Bayen, and Yi~Wu.
\newblock The surprising effectiveness of ppo in cooperative, multi-agent
  games.
\newblock \emph{arXiv preprint arXiv:2103.01955}, 2021.

\bibitem[Yu et~al.(2023)Yu, Yang, Gao, Chen, Li, Liu, Xiang, Huang, Yang, Wu,
  and Wang]{DBLP:conf/atal/YuYGCLLXHYWW23}
Chao Yu, Xinyi Yang, Jiaxuan Gao, Jiayu Chen, Yunfei Li, Jijia Liu, Yunfei
  Xiang, Ruixin Huang, Huazhong Yang, Yi~Wu, and Yu~Wang.
\newblock Asynchronous multi-agent reinforcement learning for efficient
  real-time multi-robot cooperative exploration.
\newblock In Noa Agmon, Bo~An, Alessandro Ricci, and William Yeoh (eds.),
  \emph{Proceedings of the 2023 International Conference on Autonomous Agents
  and Multiagent Systems, {AAMAS} 2023, London, United Kingdom, 29 May 2023 - 2
  June 2023}, pp.\  1107--1115. {ACM}, 2023.
\newblock \doi{10.5555/3545946.3598752}.
\newblock URL \url{https://dl.acm.org/doi/10.5555/3545946.3598752}.

\bibitem[Zhang et~al.(2022)Zhang, Tian, Zhang, Xue, Xie, Yang, Ge, and
  Chen]{zhang2022neighborhood}
Chengwei Zhang, Yu~Tian, Zhibin Zhang, Wanli Xue, Xiaofei Xie, Tianpei Yang,
  Xin Ge, and Rong Chen.
\newblock Neighborhood cooperative multiagent reinforcement learning for
  adaptive traffic signal control in epidemic regions.
\newblock \emph{IEEE Transactions on Intelligent Transportation Systems},
  23\penalty0 (12):\penalty0 25157--25168, 2022.

\bibitem[Zhou et~al.(2020)Zhou, Luo, Villella, Yang, Rusu, Miao, Zhang, Alban,
  Fadakar, Chen, et~al.]{zhou2020smarts}
Ming Zhou, Jun Luo, Julian Villella, Yaodong Yang, David Rusu, Jiayu Miao,
  Weinan Zhang, Montgomery Alban, Iman Fadakar, Zheng Chen, et~al.
\newblock Smarts: Scalable multi-agent reinforcement learning training school
  for autonomous driving.
\newblock \emph{arXiv preprint arXiv:2010.09776}, 2020.

\end{thebibliography}
\bibliographystyle{rlc}

\appendix

\section{Discussions}

\subsection{Societal Impact}
While our algorithms are performing well in SMAC and GRF, there are two major limitations. First, our algorithmic framework MAPF-LNS relies on the fact that in big and complex environments, not all agents would be used to fulfill a single task, and agents would learn to split the job into sub-tasks and do their own parts. However, the Multi-domain Gaussian Squeeze (MGS) environment used in the Qtran paper \cite{son2019qtran} is one counter-example that requires strong cooperation within all agents and thus leads to unstable and slow convergence when using our algorithms. Second, our current random-based neighborhood selection algorithms assume that all agents are similar, and no agents have a higher priority, higher importance, or significant differences from others in the environment. This is not applicable in some real-world scenarios where certain agents should have higher priority, for example, a fire engine in the traffic system. For the same reason, the marginal distribution could be ignored during training. This may lead to fairness issues if the neighborhood selection is not taking these into account (as in the proposed RLNS and BLNS), and this could lead to a potential negative social impact. We do not see any other potential negative societal impact of our work.

\subsection{Naming}
About the naming of our algorithm, we have considered using a different name for our algorithm that does not include LNS. However, two primary reasons led to our decision to keep the name with LNS: 1). Although our algorithm is not fully aligned with other LNS approaches, our algorithm is still optimizing on a local neighborhood at each iteration, which is the essence of an LNS algorithm. While the high-level idea of training on subsets of agents has been explored and called differently in different communities, e.g., subset-based optimization, random coordinate descent, and LNS, using the name of LNS could connect our future work with the rich literature in the combinatorial search community on LNS, for example, neighborhood selection. 2). We believe that our algorithm is the closest algorithm to LNS in the MARL community. The main difference between our framework and a typical LNS algorithm, which usually only changes the selected neighborhood subset, is that our MARL RLNS and BLNS do not completely fix the policy of agents that are not in the neighborhood because of parameter sharing between the policy of different agents. However, the common practice of using parameter sharing in MARL is the key to making the training efficient. Our improvement in terms of time is slightly smaller than the one brought by parameter sharing, so we believe it is necessary to keep parameter sharing in the framework.

\subsection{Theorem on Convergence}

Remark that in this paper, we are addressing the efficiency while not expecting the algorithm to outperform the low-level algorithms, we do not guarantee the number of training per LNS neighborhood is long enough for the value function and policy function to converge. Additionally, in the most recent iteration of the paper, another update of the theorem is introduced, which, diverging from previous requirements, does not necessitate the cumulative reward function to be Lipschitz differentiable within each neighborhood partition, instead necessitating a weaker inequality condition. Given that differentiable is common in the proof of convergence in numerous other Multi-Agent Reinforcement Learning (MARL) algorithms, which are highly likely to be employed as the low-level algorithm to guarantee the algorithms can learn to cooperate, we have elected not to incorporate the update here. 

\subsection{Neighborhood Selection}

While neighborhood selection is one of the most important parts of LNS research, in this paper, our neighborhood selection strategy is purely random based. There are three major reasons for this: 1. Random selection itself is very strong and robust in domains that are not well-studied by LNS researchers. Many proposed heuristics for neighborhood selection failed to outperform random selection on domains that are not included in their paper or even much worse. While the primary purpose of this paper is to introduce LNS into the context of MARL, a pure random-based algorithm variant that stably outperforms the low-level MAPPO is already good enough to fulfill the objective. 2. While MARL is a very general algorithm that could be used for many environments, how to propose a very strong and general neighborhood selection algorithm is a very big challenge. If one wants to bring the generalizability of machine learning into this domain, one might need to think about how to tradeoff the additional time used in training, such as a neighborhood selection model, and also the inference time used in the training process may reduce the total time reduction. 3. Even in the case that we do not want generalizable heuristics, we have tried in SMAC based on geographical clustering to choose local neighborhoods to align with the previous works in geographical clustering \cite{zhang2022neighborhood}. This works no better than random, and is particularly bad in the 3s5zvs3s6z scenario, whereas we discussed in the main paper, including both type of agents are necessary for each neighborhood to get a strong policy. 4. Although there are only two main branches in cooperative MARL, namely value-based and policy-based algorithms, the concrete training details are completely different from one MARL algorithm to another. And what is available for a neighborhood selection algorithm to use is also very different. For example, for MAPPO, the algorithm only has a joint value function learned, and the optimization is based on the advantage function in each episode, while in QMIX, each agent also has a local Q function that takes action into account. These differences can make a neighborhood selection algorithm like choosing the agents whose local Q function is the smallest in Qmix not applicable to MAPPO. But our current random-based algorithm does not have such a problem so we really recommend the current group of RLNS, BLNS and ALNS as a general solution if one just want to get an easy speedup, no matter what their low-level algorithm is.

\subsection{Rejecting Bad Neighborhoods}

In the current version of the paper, we do not reject any bad neighborhoods, primarily because the training of MARL always includes a lot of fluctuations caused by both local optimal and exploration, as shown in Fig.~\ref{fig:training_curve} and Fig.~\ref{fig:abla_m}. If we reject a neighborhood just by ad-hoc reduction of the evaluation win rate, the learned policy will end up in an early-point local optimal that is far from optimal. Another minor reason is that our main focus of this paper is to provide this robust MARL-LNS framework that gains the total training time reduction with nearly no extra effort, and clever rejection criteria are something that one may see as a huge burden and are what we want to avoid. However, it is undeniable that developing a clever rejecting heuristic can help the algorithms to be more stable, and we would like to leave such research in the future.

\subsection{Improvement Margin}
In this paper, our improvement is from $5\%$ to $25\%$ , depending on the scenarios and settings of our algorithms. It needs to be addressed that our reported results are obtained on a server that highly prioritizes the capacity of GPU, and thus the sampling time (which is CPU dependent) takes a larger portion of time than it would on a server that balances the configuration of CPU and the GPU. When moving to such a server that replaces the V100 with an NVIDIA P100 GPU, the improvement for 5mvs6m scenario in SMAC, which is the scenario that we have the least improvement ratio, increases from $5\%$ to $15\%$, and could reach a time saving of $25\%$ when everything is done on a single 16-core xeon-6130 CPU with 32GB memory without any GPU.

\section{Implementation Details for Experiments}
Here we provide the hyperparameter table for our experiments. While most results in the experiment section are from previous papers \cite{yu2021surprising} and \cite{wu2021coordinated}, we only provide the hyperparameter for our algorithms.

Table.~\ref{tab:hyperparam_m} provides the neighborhood size used specifically for each scenario, while Table.~\ref{tab:hyperparam_mappo} provide other hyperparameters that exist in the low-level training algorithm MAPPO. Hyperparameters are not specified by search but by directly using the recommended variables used by the low-level training algorithm MAPPO, except for the number of parallel environments which is limited by the core of our server. In SMAC environments, the number of neighborhood iterations is set to $8$.

All results displayed in this paper are trained on servers with a 16-cores xeon-6130 2.10 GHz CPU with 64GB memory, and an NVIDIA V100 24GB GPU. 

\begin{table}[h]
\centering
\begin{tabular}{@{}ll@{}}
\toprule
           & Neighborhood size $m$  \\ \midrule
5mvs6m     & 3  \\
mmm2       & 2  \\
3s5zvs3z6z & 5  \\
27mvs30m   & 15 \\
10mvs11m   & 5  \\ \bottomrule
\end{tabular}
\caption{Hyperparameter Table for neighborhood size used in RLNS and BLNS.}
\label{tab:hyperparam_m}
\end{table}

\begin{table}[h]
\centering
\begin{tabular}{@{}ll@{}}
\toprule
Hyperparameters      & value                             \\ \midrule
recurrent data chunk length & 10                                \\
gradient clip norm          & 10.0                              \\
gae lamda                   & 0.95                              \\
gamma                       & 0.99                              \\
value loss                  & huber loss                        \\
huber delta                 & 10.0                              \\
batch size num              & envs × buffer length × num agents \\
mini batch size             & batch size / mini-batch           \\
optimizer                   & Adam                              \\
optimizer epsilon           & 1e-5                              \\
weight decay                & 0                                 \\
network initialization      & Orthogonal                        \\
use reward normalization    & True                              \\
use feature normalization   & True                              \\
num envs (SMAC)             & 8                                 \\
num envs (GRF)              & 15                                \\
buffer length               & 400                               \\
num GRU layers              & 1                                 \\
RNN hidden state dim        & 64                                \\
fc layer dim                & 64                                \\
num fc                      & 2                                 \\
num fc after                & 1                                 \\ \bottomrule
\end{tabular}
\caption{Hyperparameter table for the MAPPO training part used in BLNS and RLNS.}
\label{tab:hyperparam_mappo}
\end{table}

\section{Additional Experiment Results}

\begin{figure*}[htb]
    \centering
    \begin{subfigure}[b]{0.4\textwidth}
        \centering
        \includegraphics[width=0.95\columnwidth]{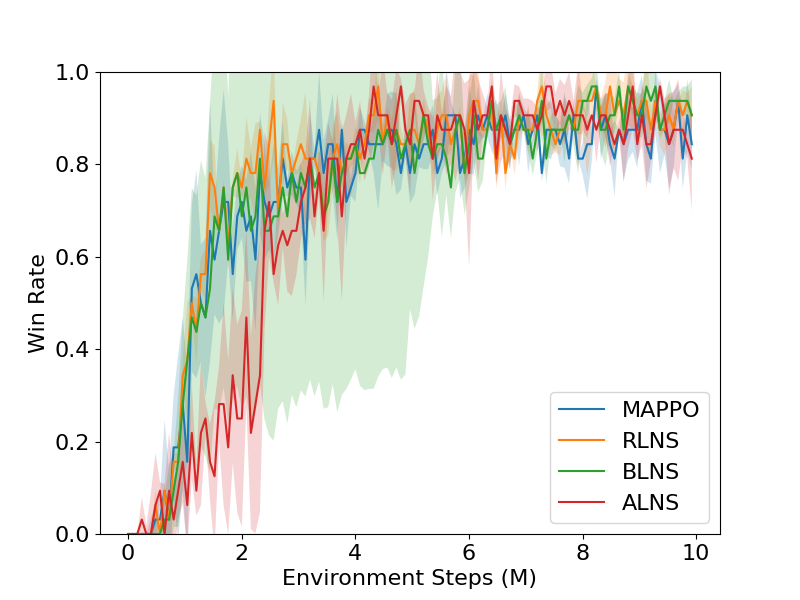}
        \caption{5M vs 6M}
        \label{fig:5m6mcurve}
    \end{subfigure}
    \centering
    \begin{subfigure}[b]{0.4\textwidth}
        \centering
        \includegraphics[width=0.95\columnwidth]{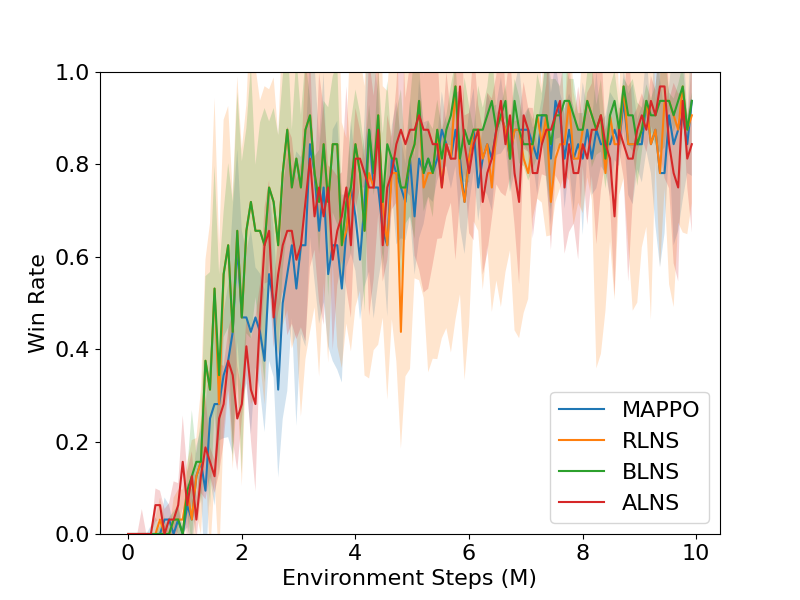}
        \caption{MMM2}
    \end{subfigure}
    \caption{Median value and standard deviation of the RLNS, BLNS, and ALNS training curves compared to MAPPO on two SMAC scenarios. Although the neighborhood size is set as half of the total number of agents, the training curves are not much different.}
    \label{fig:training_curve}
\end{figure*}

\begin{figure*}[htb]
    \centering
    \begin{subfigure}[b]{0.4\textwidth}
        \centering
        \includegraphics[width=0.85\columnwidth]{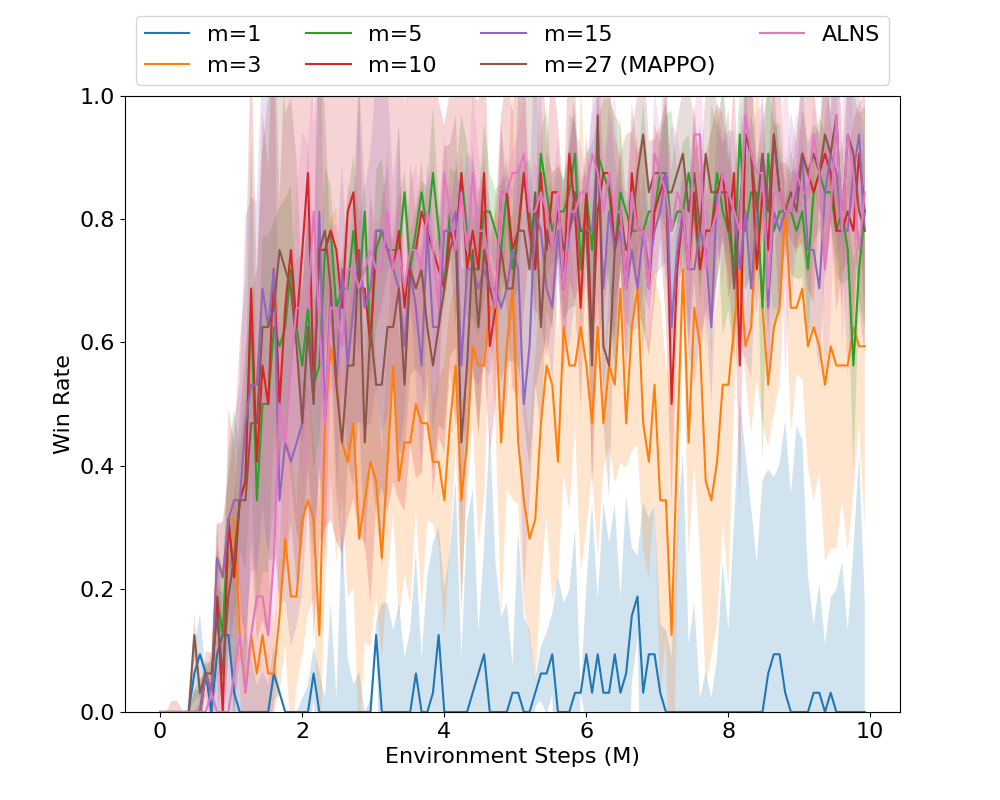}
        \caption{Win rate corresponds vs. environment step}
        \label{fig:abla_m_step}
    \end{subfigure}
    \centering
    \begin{subfigure}[b]{0.4\textwidth}
        \centering
        \includegraphics[width=0.85\columnwidth]{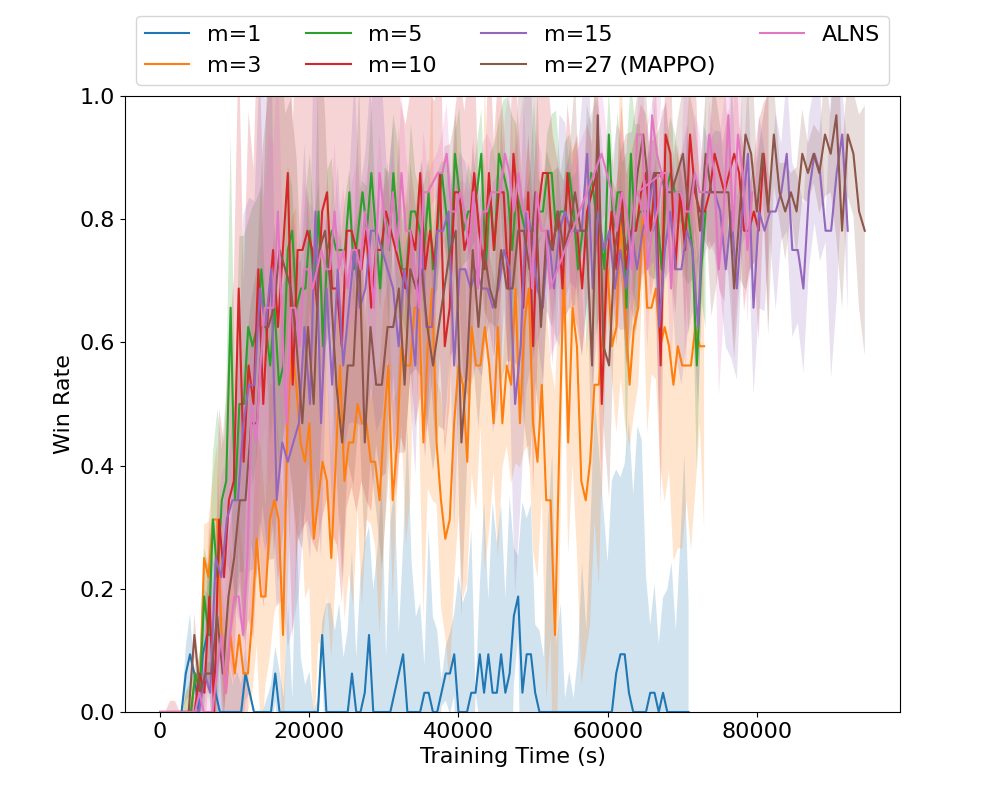}
        \caption{Win rate corresponds vs. wall clock time}
        \label{fig:abla_m_time}
    \end{subfigure}
    \caption{Median value and standard deviation of the BLNS training curve on the 27m\_vs\_30m scenario on SMAC for different neighborhood sizes $m$.}
    \label{fig:abla_m}
\end{figure*}

We have included our training curve in Fig.~\ref{fig:training_curve, fig:abla_m}. While it is acknowledgeable that the training curves have many fluctuations due to the reported values being median values, we can still see the same conclusion as we get from the table in the main paper: our proposed algorithms are as good as our base-level algorithm MAPPO. A reasonable neighborhood size $m$ that is larger than $3$ can also make our algorithm not significantly worse than the base algorithm.

Besides, as shown in Fig.~\ref{fig:5m6mcurve}, we observe that ALNS may not always be the most efficient at any time, given that training on a limited number of agents like 2 may lead to a slow improvement on policy function, but ALNS is as good as the low-level algorithm as the training progress and ALNS increase the neighborhood size.

\section{Additional Theoretical Guarantee for Convergence of MARL-LNS}

In the main paper, we have provided the theorem that guarantees the convergence of MARL-LNS to be the same as the low-level algorithm. Here we provide a stronger theorem in the case that in each LNS iteration, the policy is updated to local optimal.

\begin{theorem}
\label{thm:converge_exact}
(Adapted from \cite{lyu2020convergence}) Assume the expected cumulative reward function $\mathcal{J}$ is continuously differentiable with Lipschitz gradient and convex in each neighborhood partition, and the training by the low-level algorithm guarantees that the training happening on the i-th neighborhood is bounded by some high-dimension vector $w_i$. Suppose $\sum_{i=1}^{\infty} w_i^2 < \infty$, then the following hold:
\begin{enumerate}
    \item If $\sum_{i=1}^{\infty} |w_i| = \infty$, then for any initial starting point of training, the training can converge to a stationary point.
    \item If $\pi^k$ is optimized to optimal in each LNS iteration, then there exists some constant $c > 0$ such that for $i \ge 1$, 
    \begin{align}
    min_{1\le k \le i}  \sup_{\pi_0 \in \Pi}[-\inf_{\pi \in \Pi} \nonumber \langle \nabla \mathcal{J}_{\pi^{k}}, \frac{\pi-\pi^{k}}{|\pi-\pi^{k}|}\rangle] \le \frac{c}{\sum_{k=1}^i w_k}    
    \end{align}
\end{enumerate}
\end{theorem}

Compared to the one used in the main paper, this theorem relies on the assumption of the optimality gap to be summable. However in general practice, always making the policy optimized to optimal in each LNS iteration will lead to an extremely long training time and, thus, is less preferable.

\end{document}